\documentclass[12pt]{article}

\pdfoutput=1

\usepackage{putex}
\usepackage{graphicx}

\begin{document}

\preprint{PUPT-2242 \\ LMU-ASC 63/07}

\institution{PU}{Joseph Henry Laboratories, Princeton University, Princeton, NJ 08544, USA}
\institution{MaxPlanck}{Ludwig-Maximilians-Universit\"at, Department f\"ur Physik, Theresienstrasse 37, \cr 80333 M\"unchen, Germany}

\title{Universality of the diffusion wake in the gauge-string duality}

\authors{Steven S. Gubser\worksat{\PU,}\footnote{e-mail: {\tt ssgubser@Princeton.EDU}} and
Amos Yarom\worksat{\MaxPlanck,}\footnote{e-mail: {\tt yarom@theorie.physik.uni-muenchen.de}}}

\abstract{As a particle moves through a fluid, it may generate a laminar wake behind it.  In the gauge-string duality, we show that such a diffusion wake is created by a heavy quark moving through a thermal plasma and that it has a universal strength when compared to the total drag force exerted on the quark by the plasma.  The universality extends over all asymptotically anti-de Sitter supergravity constructions with arbitrary scalar matter.  We discuss how these results relate to the linearized hydrodynamic approximation
and how they bear on our understanding of di-hadron correlators in heavy ion collisions.}

\date{September 2007}

\maketitle

\tableofcontents

\section{Introduction}
\label{INTRODUCTION}

\begin{figure}
  \begin{center}
   \includegraphics[width=4in]{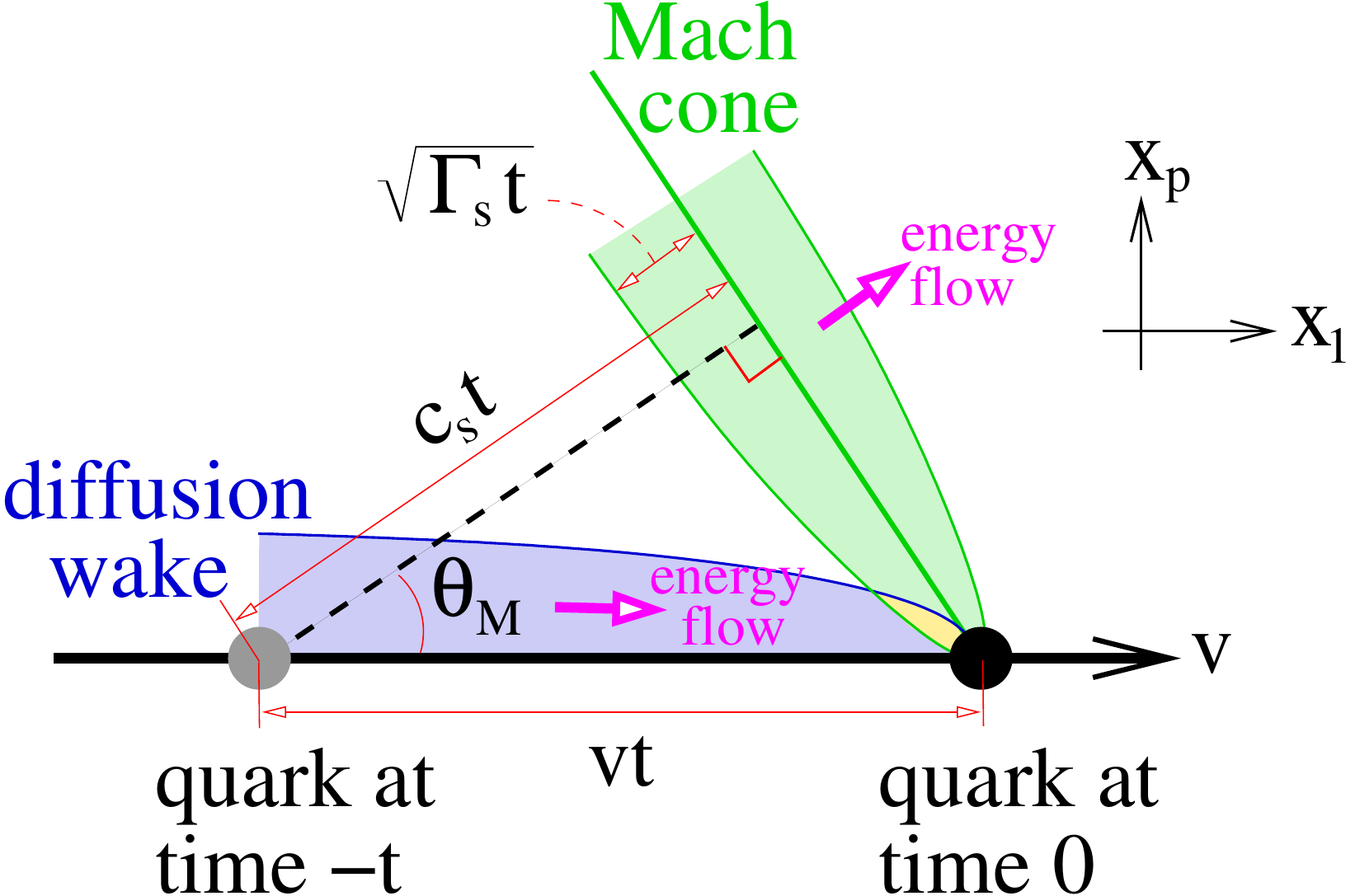}
   \vskip0.5in
   \includegraphics[width=4in]{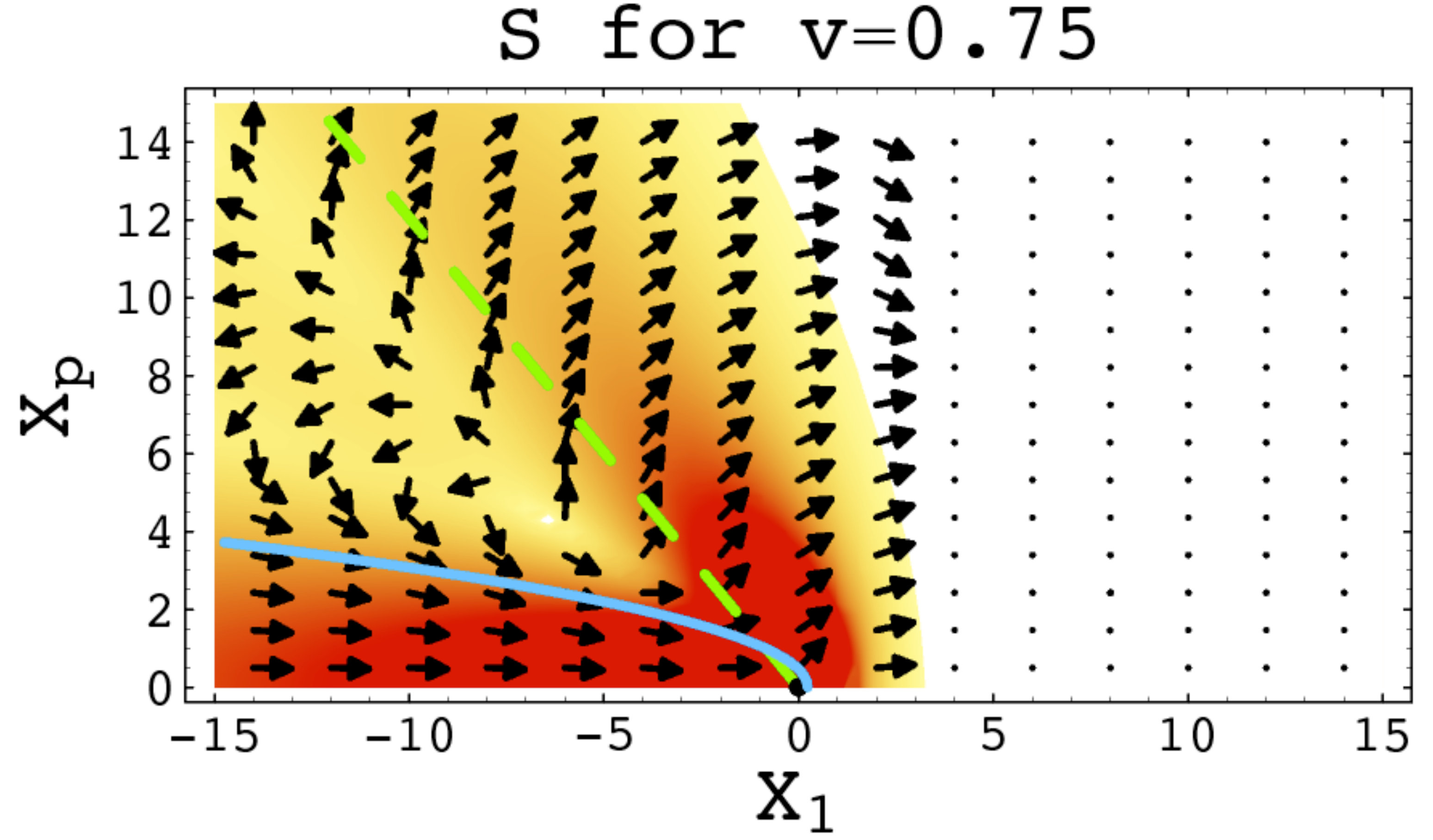}
  \end{center}
  \caption{Top: A cartoon of the diffusion wake and the sonic boom.  The Mach cone and the diffusion wake experience comparable viscosity broadening, controlled by the parameter $\Gamma_s = 4\eta/3Ts$.  Bottom: The Poynting vector $\vec{S}$ produced by a heavy quark propagating through a thermal plasma of ${\cal N}=4$ super-Yang-Mills theory, from \cite{Gubser:2007ga}.  The arrows indicate the direction of $\vec{S}$, and the color indicates its magnitude, with red meaning large $|\vec{S}|$. The dashed green line marks the Mach angle, and the gray line marks the curve where the far field asymptotics of the wake falls to half its maximal value.}\label{WakeAndBoom}
 \end{figure}

Consider a quark moving at constant velocity through an infinite, static, thermal medium.  Assume that the temperature is above the confinement transition, so the medium is a plasma of gluons and whatever dynamical matter there may be.  The quark loses energy to the plasma, but where does this energy go?  Sufficiently far from the quark, a complete account of energy dissipation should be possible within the framework of linearized hydrodynamics. Hydrodynamics is usually applicable at length scales much larger than the mean free path of thermal quasi-particles within the plasma, which is to say some multiple of $1/T$.  (We set $\hbar = c = k_B = 1$ throughout.)  The applicability of linearized hydrodynamics depends on being far enough from the quark for the perturbations to be very small. It is well understood (see for example \cite{Casalderrey-Solana:2004qm}) that if one assumes axial symmetry of the fluid flow around the direction of motion of the quark, then the equations of linearized hydrodynamics partially decouple into a pair of equations describing compressive waves (sound) and a third equation describing shearing modes (the diffusion wake).\footnote{In principle, there could be additional modes coming from conserved currents. In the scenarios we consider, such currents do not appear.}  Figure~\ref{WakeAndBoom} shows a typical geometry for these two modes.

We denote the total energy of the moving quark as $E$.  The total energy gained by the plasma is given by $P^{\rm t} = - dE/dt$, where in our conventions $dE/dt < 0$ means that energy is flowing from the quark into the plasma.   Assuming that sound and diffusion are the only modes of energy loss from the quark, the total power delivered to the plasma is
 \eqn{EnergySum}{
  P^{\rm t} = P^{\rm d} +
    P^{\rm s}\,,
 }
where $P^{\rm s}$ and $P^{\rm d}$ correspond respectively to the energy lost to sound and the energy lost to diffusion.

We claim that in a fairly broad range of string theory constructions and using a sensible but non-unique way of separating diffusion modes from other modes,
 \eqn{DiffusionRatio}{
  {P^{\rm d}} :
    {P^{\rm t}} = -1 : v^2 \,.
 }
The relative minus sign means that the diffusion wake supplies energy to the quark.  We prove this in section~\ref{DIFFUSION} for any theory admitting a holographic dual asymptotic to AdS${}_5$ that can be described in terms of Einstein gravity coupled to scalars with no more than two derivatives in the action.  In our demonstration, the quark has to be infinitely massive, and it is represented as a string trailing into the five-dimensional dual geometry.\footnote{Clearly an infinitely massive quark must also have divergent energy $E$.  But $P^{\rm t} = -dE/dt$ is finite and closely related to the drag force as calculated for ${\cal N}=4$ super-Yang-Mills theory in \cite{Herzog:2006gh,Gubser:2006bz}; see also the discussion at the end of section~\ref{TRAILING}.}  The tension of the string is an arbitrary function of the scalars.

Putting \eno{EnergySum} and \eno{DiffusionRatio} together, we learn that
\eqn{EnergyRatio}{
  {P^{\rm s}} : {P^{\rm d}} :
    {P^{\rm t}} = 1+v^2 : -1 : v^2 \,.
 }
This means that energy flows into sound waves at a rate which is $1+1/v^2$ times as fast as the total outflow of energy from the quark.  The diffusion wake supplies energy to the quark in just the right amount to balance equation \eno{EnergySum}.  In section~\ref{SOUND} we show that we do not need to assume \eno{EnergySum} in order to demonstrate \eno{EnergyRatio}, at least in the case of a single scalar.  The result \eno{EnergyRatio} holds for all $v$ between $0$ and $1$, but the most interesting case is when $v$ is greater than the speed of sound $c_s$, in which case there is a sonic boom.

The result \eno{EnergyRatio} was demonstrated for ${\cal N}=4$ super-Yang-Mills with a large number of colors and a large 't~Hooft coupling in \cite{Gubser:2007ga}, which followed upon a number of works by ourselves and other authors \cite{Herzog:2006gh,Gubser:2006bz,Friess:2006aw,Yarom:2007ap,Gubser:2007nd,Gubser:2007xz,Chesler:2007an}; parallel lines of development have included \cite{Sin:2004yx,Liu:2006ug,Casalderrey-Solana:2006rq,Sin:2006yz,Liu:2006he,Gubser:2006nz,Casalderrey-Solana:2007qw,Lin:2007pv}.  We note in particular the close relation of \cite{Casalderrey-Solana:2006rq} to the drag force results of \cite{Herzog:2006gh,Gubser:2006bz}.

The backgrounds we consider can accommodate arbitrary speed of sound (limited, perhaps, by $c_s \leq 1/\sqrt{3}$), including a speed of sound that varies as a function of temperature.  Thus our result \eno{EnergyRatio} has a degree of universality which encourages us to think that it might be compared with data from heavy ion collisions.  As we explain in section~\ref{DISCUSSION}, a diffusion wake as strong as we find poses a challenge to our understanding of the experimental situation.

\section{The diffusion wake}
\label{DIFFUSION}

The action we start with is
 \eqn{GravityAction}{
  S &= {1 \over 2\kappa_5^2} \int_M d^5 x \, \sqrt{-G}
    \left( R + L_\phi  \right) -
   {1 \over 2\pi\alpha'} \int_W d^2 \sigma \, \sqrt{-g} \, q(\vec\phi)
    \,,
 }
where $L_\phi$ is the scalar Lagrangian
\begin{equation}
\label{E:matterL}
    L_\phi = - {1 \over 2} G^{\mu\nu}
      \Omega_{IJ}(\vec{\phi})\partial_\mu \phi^I
        \partial_\nu \phi^J - V(\vec{\phi}) \,.
\end{equation}
$G_{\mu\nu}$ is the metric on the bulk spacetime $M$, and $g_{\alpha\beta}$ is the metric induced from it on the string worldsheet $W$.  We will consider backgrounds which are asymptotically anti-de Sitter, with $\vec\phi \to \vec\phi_B$ near the conformal boundary of AdS. The functions $V(\phi)$, $\Omega(\vec{\phi})$ and $q(\vec{\phi})$ are arbitrary, except for the requirement that $V(\phi)$ should have a local (negative) maximum for a finite value of the scalars, $V(\vec{\phi}_B)=-12/L^2$, in order for the desired anti-de Sitter asymptotics to arise near the conformal boundary.

When the string is not present, the backgrounds of interest take the form
 \eqn{BHansatz}{
  ds^2 = \alpha(r)^2 \left[ -h(r) dt^2 + d\vec{x}^2 +
     {dr^2 \over h(r)} \right] \qquad \vec{\phi} = \vec{\phi}(r) \,.
 }
This ansatz is the result of assuming translation invariance in the ${\bf R}^{3,1}$ directions and rotational invariance in the ${\bf R}^3$ directions.
It is assumed that $h \to 1$ at the conformal boundary where also $r \to 0$ and $\alpha \to L/r$.  There is an event horizon at $r = r_H > 0$ where $h(r_H)=0$.  The $tt$ and $ii$ components of the Einstein equations (where $i$ runs over the three components of $\vec{x}$) provide us with
\begin{subequations}
\label{BHeoms}
\begin{align}
\label{E:BHeomsa}
    \frac{d}{dr}\left(h^{\prime}\alpha^3\right) &= 0\\
\label{E:BHeomsb}
    \frac{3}{2}\frac{h^{\prime}\alpha^{\prime}}{h\alpha} + 3\frac{\alpha^{\prime\prime}}{\alpha} &= \frac{1}{2}\frac{\alpha^2}{h}L_{\phi}\,,
\end{align}
where primes denote $d/dr$.  The scalars satisfy
\begin{equation}
\label{E:BHeomsc}
    {d \over dr} \frac{\partial L_\phi}{\partial \phi^{I\,\prime}} - \frac{\partial L_\phi}{\partial \phi^I} = 0 \,.
\end{equation}
\end{subequations}
In addition, the $rr$ component of the Einstein equation imposes a zero-energy constraint:
 \eqn{BHconstraint}{
        \frac{3}{2}\frac{h^{\prime}\alpha^{\prime}}{h\alpha}
        +6\frac{\alpha^{\prime\,2}}{\alpha^2}=
    \frac{1}{2}\frac{\alpha^2}{h}\left( L_\phi + \Omega_{IJ}\phi^{I\,\prime}\phi^{J\,\prime}\right) \,.
 }

The system of equations \eno{BHeoms} and \eno{BHconstraint} has been extensively studied, and we briefly summarize aspects of the relevant literature.  It was shown in \cite{Gubser:2000nd} that a solution of the form \eno{BHansatz} can exist in the case of a single scalar only if $V(\phi_H) < V(\phi_B)$ with $\phi_H \equiv \phi(r_H)$.  Numerical construction of such black holes was first reported on in \cite{Gubser:2000nd} for a specific potential.  Further numerical solutions were exhibited more explicitly in \cite{Friess:2005zp} and later in \cite{PufuUnpublished,Buchel:2007vy,PandoZayas:2006sa,Aharony:2007vg,Mahato:2007zm}, where analyses of thermodynamic quantities were also given.  Following \cite{Buchel:2000ch}, some approximate, analytical results for high temperature black holes were obtained in \cite{Gubser:2001ri,Buchel:2001gw,Buchel:2003ah}.

\subsection{The trailing string}
\label{TRAILING}

In \cite{Herzog:2006gh,Gubser:2006bz} the drag force acting on a moving quark at constant velocity $v$ through a plasma has been calculated using AdS/CFT. It is straightforward to generalize the drag force calculation of \cite{Herzog:2006gh,Gubser:2006bz} to the backgrounds described above.  Various generalizations exist in the literature, for example \cite{Herzog:2006se,Talavera:2006tj,MichalogiorgakisPhD}.  The one we will describe here is quite a modest extension of previously published results, which allows for an arbitrary dependence on the scalar field $\phi^I$  coming from a Kaluza-Klein reduction of the ten dimensional Dilaton and metric.  The ansatz for the trailing string is
 \eqn{AnsatzTrailing}{
  x^1 = vt + \xi(r) \,,
 }
and the string action takes the form
 \eqn{Strailing}{
  S = \int dt dr \, L \qquad
   L = -{q(\vec{\phi}) \over 2\pi\alpha'}
    \alpha^2 \sqrt{1 - {v^2 \over h} + h \xi'^2} \,.
 }
The conserved momentum conjugate to $\xi$ is
 \eqn{pixiDef}{
  \pi_\xi = {\partial L \over \partial\xi'} =
  -{q(\vec{\phi}) \over 2\pi\alpha'} {h\alpha^2 \over
   \sqrt{1 - v^2/h + h \xi'^2}} \xi' \,.
 }
The drag force exerted by the plasma on the quark is
 \eqn{DragForce}{
  F_1 = {dp_1 \over dt} = -\pi_\xi \,.
 }
Note that $\pi_\xi > 0$, so $F_1$ is negative, meaning in a direction opposite the quark's velocity.  The total rate of energy loss from the quark to the plasma is
 \eqn{TotalEnergyLoss}{
  P^{\rm t} = -{dE \over dt} = -v {dE \over dx} =
    -v {dp_1 \over dt} = v\pi_\xi \,,
 }
which is a positive quantity.

\subsection{Perturbations in axial gauge}
\label{AXIAL}

The trailing string sources the metric and the scalar fields coupled to it.  Treating $\kappa_5^2$ as a small parameter, we may evaluate the response of the various fields to the string to linear order in $\kappa_5^2$.

We consider linearized fluctuations around a background of the form \eqref{BHansatz}.
These fluctuations will be expressed as
 \eqn{LinearizedFluctuations}{
  G_{\mu\nu} \to G_{\mu\nu} + \delta G_{\mu\nu}, \qquad
   \vec{\phi} \to \vec{\phi} + \delta\vec{\phi}, \qquad \hbox{etc.}
 }
In axial gauge one sets $\delta G_{r\nu} = 0$ for all $\nu$. We further define
 \eqn{deltaGH}{
  \delta G_{mn} = \kappa_5^2 \alpha^2 H_{mn} \,,
 }
where the pre-factor $\kappa_5^2\alpha^2$ is for convenience.

As in \cite{Friess:2006fk}, we study the linearized fluctuations in Fourier space with a co-moving ansatz:
 \eqn{HmnFourier}{
  H_{mn}(t,r,\vec{x}) = \int {d^3 k \over (2\pi)^3}
    e^{i k_1 (x^1-vt) + k_2 x^2 + k_3 x^3} H_{mn}(t,r,\vec{k}) \,.
 }
Because of axial symmetry, the Fourier coefficients $H_{mn}$ can only depend on wave-number $\vec{k}$ through $k_1$ and $k_\perp \equiv \sqrt{k_2^2 + k_3^2}$.  Let's fix this symmetry by setting $k_3=0$ and $k_2=k_\perp>0$.  The three components ($H_{03}$, $H_{13}$, and $H_{23}$) are odd under reflection through the plane perpendicular to $\hat{3}$.  The trailing string is symmetric under this reflection, so it doesn't source the odd metric components.  We may therefore set them to zero.  The other components ($H_{00}$, $H_{01}$, $H_{02}$, $H_{11}$, $H_{12}$, $H_{22}$, and $H_{33}$) are even, and they are all sourced non-trivially by the trailing string.
A substantial simplification of the equations of motion arises from forming the following linear combinations, which are essentially the same as the ones in \cite{Friess:2006fk}:
\begin{align}
\label{E:defA}
	A &= \frac{1}{2v^2}\left(-H_{11}+2\frac{k_1}{k_{\bot}} H_{12} - \left(\frac{k_1}{k_{\bot}}\right)^2 H_{22}
		+ \left(\frac{k}{k_{\bot}}\right)^2 H_{33} \right)
\end{align}
 \eqn{Drul}{	
	\vec{D} &= \begin{pmatrix}
		D_1 \\ D_2
		  \end{pmatrix}
	  	= \begin{pmatrix}
	  	 \frac{1}{2v}\left(H_{01}-\frac{k_1}{k_{\bot}} H_{02}\right)\\
		  \frac{1}{2v^2}
		  \left(-H_{11} + \left(\frac{k_1}{k_{\bot}}-\frac{k_\bot}{k_1}\right) H_{12}
		  +H_{22} \right)
		 \end{pmatrix}
 }
\begin{equation}
\label{E:defE}
	\vec{E} = \begin{pmatrix}
		E_1 \\ E_2 \\ E_3 \\ E_4
		\end{pmatrix}
	 = \begin{pmatrix}
	 	\frac{1}{2}\left(-\frac{3}{h} H_{00}+H_{11}+H_{22}+H_{33}\right)\\
		\frac{1}{2v}\left(H_{01}+\frac{k_{\bot}}{k_1} H_{02}\right)\\
		\frac{1}{2}\left( H_{11}+H_{22}+H_{33}\right)\\
		\frac{1}{4}\left(H_{11}\left(2-6\left(\frac{k_1}{k}\right)^2\right)
			+H_{22}\left(-4+6\left(\frac{k_1}{k}\right)^2\right)
			+2 H_{33} - 12 \frac{k_1 k_{\bot}}{k^2} H_{12}\right)
	    \end{pmatrix} \,.
\end{equation}
The equation of motion for $A$ decouples from all other metric components as well as from the scalars.  The equations of motion for $\vec{D}$ mix with one another but decouple from $A$ and $\vec{E}$ as well as the scalars.  The equations of motion for $\vec{E}$ couple to the scalars.  The defining relations \eno{E:defA}-\eno{E:defE} are equivalent to
 \eqn{Hrules}{
  H_{00} &= {2h \over 3} (-E_1 + E_3)  \cr
  H_{01} &= {2v \over k^2} (k_\perp^2 D_1 + k_1^2 E_2)  \cr
  H_{02} &= {2v k_1 k_\perp \over k^2} (-D_1 + E_2)  \cr
  H_{11} &= -{v^2 k_\perp^4 \over k^4} A -
    {4v^2 k_1^2 k_\perp^2 \over k^4} D_2 + {2 \over 3} E_3 +
    {-2k_1^2 + k_\perp^2 \over 3k^2} E_4  \cr
  H_{12} &= {k_1 k_\perp \over k^2} \left(
    {v^2 k_\perp^2 \over k^2} A +
    2v^2 {k_1^2 - k_\perp^2 \over k^2} D_2 - E_4 \right)  \cr
  H_{22} &= -{v^2 k_1^2 k_\perp^2 \over k^4} A +
    {4v^2 k_1^2 k_\perp^2 D_2 \over k^4} +
    {2 \over 3} E_3 +
    {k_1^2 - 2k_\perp^2 \over 3k^2} E_4  \cr
  H_{33} &= {v^2 k_\perp^2 \over k^2} A + {2 \over 3} E_3 +
    {1 \over 3} E_4 \,.
 }
It will prove useful to have these inverse relations ready to hand when we want to extract components of the gauge theory's stress energy tensor.

We shall restrict our attention to the vector modes since these are the only ones which will contribute to the diffusion wake. To see this consider the contribution of the moving quark to the Poynting vector
 \eqn{PoyntingDef}{
  \vec{S} = \begin{pmatrix} S_1 & S_2 & S_3 \end{pmatrix}
    \qquad\hbox{where} \quad
    S_i = \langle \delta T^{0i} \rangle = -\langle
     \delta T_{0i} \rangle \,.
 }
Recall that we employ mostly plus signature and that $\langle T^{0i} \rangle = 0$ in the static background.  We use $\vec{S}$ in preference to $\delta\vec{S}$ for simplicity, keeping in mind that the non-zero Poynting vector owes wholly to the presence of the moving quark.
Using holographic renormalization (see appendix~\ref{HOLORG}) we can show that
 \eqn{HolographicPoynting}{
  S_i =  - \lim_{r \to 0} {2\alpha^4 \over L} H_{0i} \,.
 }
It is useful to decompose the Fourier components of the Poynting vector into a longitudinal part and a transverse part
\begin{equation}
\label{E:separateS}
    \vec{S} = S_L \hat{k} + \vec{S}_T \,,
\end{equation}
where $\hat{k} = \frac{1}{k}(k_1,k_{\bot})$. The $\vec{S}_T$ component is orthogonal to $\hat{k}$ and lies in the same plane as $\vec{k}$ and $\vec{v}$ due to the azimuthal symmetry around the direction of motion of the quark. It is well understood (see for example \cite{Casalderrey-Solana:2004qm}) that the diffusion wake is described in terms of $\vec{S}_T$. Referring to \eno{Hrules}, we see that the contribution of $E_2$ to $\vec{S}$ is in a direction parallel to the wave-number $\vec{k}$, whereas the contribution to $D_1$ is orthogonal to $\vec{k}$. In short, $E_2$ controls $S_L$ while $D_1$ controls $\vec{S}_T$.  Using \eqref{Hrules} and \eno{HolographicPoynting} one finds
\begin{equation}
    \vec{S}_T = - {4v k_\perp^2 \over L k^2}
      \begin{pmatrix} 1 & -{k_1 \over k_\perp} & 0 \end{pmatrix}
      \lim_{r \to 0} \alpha^4 D_1\,.
\end{equation}

The linearized Einstein equations in axial gauge imply the following second order equations for the $D$ modes,
 \eqn{Deqns}{
    \alpha^{-3}
      {d \over dr} \left(D_1^{\prime}\alpha^3\right)-\frac{k^2}{h}D_1 + \frac{k_1^2 v^2 }{h} D_2 & = \frac{\pi_{\xi}}{h^2 \alpha^3 \xi^{\prime}}e^{-i k_1 \xi}  \cr
    \alpha^{-3}h^{-1} {d \over dr} \left(D_2^{\prime}h\alpha^3\right)-\frac{k^2}{h^2} D_1 + \frac{k_1^2 v^2}{h^2} & = \frac{v^2 - h^2 \xi^{\prime\,2}}{v^2 h^3 \alpha^3 \xi^{\prime}}\pi_{\xi}e^{-i k_1 \xi}\,.
 }
In \eqref{Deqns}, the real-valued function $\xi(r)$ is determined by solving \eqref{pixiDef}.  Additionally there is a first order constraint,
 \eqn{Dconstraint}{
    D_1^{\prime} - h D_2^{\prime} = \frac{\pi_{\xi}}{i k_1 v^2 \alpha^3} e^{-i k_1 \xi} \,.
 }
Note that the source for the constraint equations \eqref{Dconstraint} is of order $\mathcal{O}(k^{-1})$ while that of the second order equations \eqref{Deqns} is of order $\mathcal{O}(1)$.

\subsection{Large distance asymptotics}
\label{SMALLKDIFFUSE}

We are interested in the large distance asymptotics of the energy momentum tensor. In this regime one would expect that linearized hydrodynamics will be a good approximation. The equations for vector perturbations simplify drastically when only the leading order terms in small $k$ are retained: the equations of motion \eno{Deqns} become
 \eqn{DeomsSmallK}{
  {d \over dr} (\alpha^3 D_1') = 0  \qquad
  {d \over dr} (h\alpha^3 D_2') = 0 \,,
 }
and the first order constraint \eno{Dconstraint} becomes
 \eqn{DconstraintSmallK}{
  D_1' - h D_2' = {\pi_\xi \over i v^2 k_1 \alpha^3} \,.
 }
The most general solution of the equations of motion \eno{DeomsSmallK} consistent with the boundary condition $D_i \to 0$ at the conformal boundary of the asymptotically AdS background is
 \eqn{DsolutionsSmallK}{
  D_1(r) = d_1 \int_0^r {d\tilde{r} \over \alpha(\tilde{r})^3} \qquad
  D_2(r) = d_2 \int_0^r {d\tilde{r} \over h(\tilde{r})
    \alpha(\tilde{r})^3} \,.
 }
Because $h \to 0$ at the horizon, $D_2$ has a logarithmic divergence there.  Requiring that the metric fluctuations are finite at the horizon implies $d_2=0$.  This is equivalent to taking the $k \to 0$ limit of the usual horizon boundary condition where outgoing modes are forbidden.  A more detailed discussion of the horizon boundary condition can be found in section~\ref{SOUND}.  Plugging \eno{DsolutionsSmallK} into \eno{DconstraintSmallK} with $d_2=0$ leads immediately to
 \eqn{GotDone}{
  d_1 = {\pi_\xi \over i v^2 k_1} \,.
 }

The diffusion wake is a long, narrow, forward-directed stream of fluid along the negative $x^1$ axis (see figure~\ref{WakeAndBoom}).  Its narrowness in the $x^2$ and $x^3$ directions means that, in Fourier space, we can take $k_2$ and $k_3$ large compared to $k_1$, effectively setting $k_\perp = k$. In this limit, we find a simple limiting form for $\vec{S}_T$:
 \eqn{SimpleWake}{
   \vec{S}_T \to
     -\frac{\pi_{\xi}}{i v k_1}
     \begin{pmatrix} 1 & 0 & 0 \end{pmatrix}
     \equiv \vec{S}^{\rm d} \,.
 }
To Fourier transform $\vec{S}^{\rm d}$ to real space, we need a prescription for passing the $k_1=0$ pole.  Passing it in the upper half plane results in a stream of energy along the negative $x^1$ axis, while passing it in the lower half plane results in a stream of energy along the positive $x^1$ axis---ahead of the quark.  This latter possibility is unphysical in a steady-state solution. We find
 \eqn{NarrowWake}{
  S_1^{\rm d}(0,\vec{x}) &=
   -\frac{\pi_{\xi}}{i v}\int {d^3 k \over (2\pi)^3} e^{i \vec{k} \cdot \vec{x}}
     {1 \over k_1 + i\epsilon} \cr
    &= \frac{\pi_{\xi}}{v} \, \theta(-x^1) \delta(x^2)
        \delta(x^3) \,,
 }
where the integration is along real values, $\epsilon > 0$, and
 \eqn{DefineStepFunction}{
  \theta(x) \equiv \left\{ \seqalign{\span\TL &\qquad\span\TT}{
    1 & for $x>0$  \cr\noalign{\vskip-0.5\jot}
    0 & for $x<0$.
   } \right.
 }
In \cite{Gubser:2007ga,Gubser:2007xz}, it was observed that for the case of ${\cal N}=4$ super-Yang-Mills theory, effects of order $\mathcal{O}(1)$ lead to extra terms which correspond to replacing $\epsilon$ in \eno{NarrowWake} by ${\eta/s \over vT} k^2$.  This replacement broadens the wake from a singular structure along the negative $x_1$ axis to a stream inside a parabolic surface of revolution, as shown in figure~\ref{WakeAndBoom}.

\begin{figure}
  \centerline{\includegraphics[width=4in]{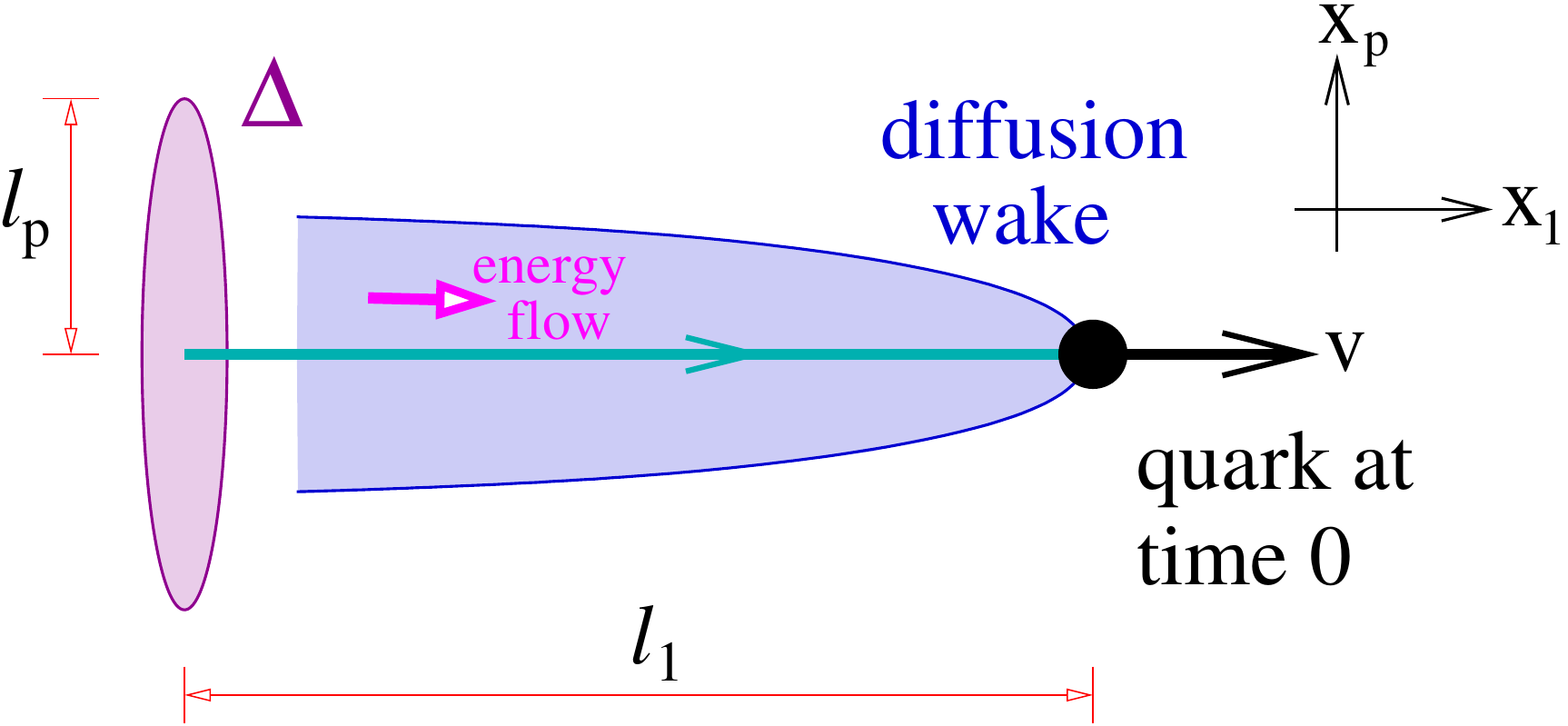}}
  \caption{The diffusion wake and the integration surface $\Delta$ used in equation \eno{DiffusionPower}.  The inviscid limit of the diffusion wake is an infinitesmially narrow, forward directed stream of energy, as shown in lighter blue.  After viscosity broadening, the diffusion wake thickens to a parabolic shape.}\label{disk}
 \end{figure}

The energy carried by the narrow stream described in \eno{NarrowWake} per unit time is given by
 \eqn{DiffusionPower}{
  {P^{\rm d}} = \oint \vec{S}^{\rm d}\cdot d\vec{a} = \int_\Delta \vec{S} \cdot d\vec{a} = -\frac{\pi_{\xi}}{v} \,.
 }
The first surface integral in \eqref{DiffusionPower} is over the boundary of a large volume enclosing the moving quark, and $d\vec{a}$ is an outward pointing area element. This integral can be evaluated immediately, using \eno{NarrowWake}, to give the result $-\pi_\xi/v$.  In writing the second integral, we aim to reproduce the same result by integrating the full Poynting vector $\vec{S}$ over a carefully chosen surface $\Delta$.  From \eno{NarrowWake} we might expect that $\Delta$ can be any sufficiently small surface intersecting the negative real axis.  But in the full Poynting vector, there is viscosity broadening, and there are also non-hydrodynamical effects close to the quark \cite{Yarom:2007ap,Gubser:2007nd,Yarom:2007ni}.  A good choice of $\Delta$ is shown in figure~\ref{disk}, where one takes $\ell_1 \to \infty$ with $\ell_1 \gg \ell_p \gg \sqrt{\ell_1 \Gamma_s}$.  The large $\ell_1$ limit insures that linearized hydrodynamics is valid, and the specified range of $\ell_p$ insures that the integration over $\Delta$ picks up the entire contribution from the diffusion wake and nothing else.  The area element $d\vec{a}$ points toward negative $x_1$ to agree with directionality of $d\vec{a}$ in the first integral in \eno{DiffusionPower}.
Combining \eqref{TotalEnergyLoss}, \eno{GotDone}, and \eno{DiffusionPower}, we have
 \eqn{FinalDP}{
  {P^{\rm d}} =
    -{1 \over v^2} P^{\rm t} \,,
 }
This is the result we advertised in \eno{DiffusionRatio}.  In section~\ref{NONCONSERVATION} we will describe another way of evaluating $P^{\rm d}$ starting from the divergence of the stress tensor.

\section{Sound modes}
\label{SOUND}

To describe the sound modes, we need to know the first column of the stress energy tensor in the small momentum limit.\footnote{Near the quark, the flow is not hydrodynamical.  A near-field account can probably be given, along the lines of \cite{Yarom:2007ap,Gubser:2007nd,Yarom:2007ni}.}  For purposes of simplicity, we restrict attention to a single scalar, so that
 \eqn{LsSimple}{
  L_\phi = -{1 \over 2} G^{\mu\nu} \partial_\mu \phi \partial_\nu \phi -
    V(\phi) \,.
 }
Referring to \eno{Hrules}, we see that to determine the Poynting vector, it is enough to understand the asymptotics of $E_2$.  This field couples non-trivially both to the scalar and to $E_1$, $E_3$, and $E_4$.  However, in the small momentum approximation a simplification occurs: two of the equations of motion decouple, giving
\begin{align}
\label{E:E2}
    \frac{d}{dr}\left(E_2^{\prime}\alpha^3\right) &= 0\\
\label{E:E4}
    \frac{d}{dr}\left(E_4^{\prime}h\alpha^3\right) &= 0 \,.
\end{align}
The other equations of motion, also at order $\mathcal{O}(1/k)$, are
\begin{subequations}
\label{EEStatic}
\begin{align}
  \delta\phi'' + \left( {3\alpha' \over \alpha} +
     {h' \over h} \right) \delta\phi' -
    {\alpha^2 V''(\phi) \over h} \delta\phi +
    \left( {1 \over 3} E_1' + {2 \over 3} E_3' \right) \phi' &= 0  \\
  E_1'' + \left( {3\alpha' \over \alpha} + {3h' \over 2h} \right) E_1' +
    \phi' \delta\phi' + {\alpha^2 V'(\phi) \over h} \delta\phi &= 0  \\
  E_3'' + \left( {3\alpha' \over \alpha} + {h' \over 2h} \right) E_3' +
    {1 \over 2} \phi' \delta\phi' +
     {\alpha^2 V'(\phi) \over 2h} \delta\phi &= 0 \,.
\end{align}
\end{subequations}
The first order constraints couple together all the $E_i$'s and the scalar:
\begin{subequations}
\label{E:ExtraConstraints}
\begin{align}
\label{E:Constraint1}
    E_2^{\prime}+E_3^{\prime}
    -\frac{h^{\prime}}{h} E_2 - \frac{h^{\prime}}{2 h} E_3 +\frac{1}{2}\phi^{\prime}\delta\phi & =
        \frac{i \pi_{\xi}}{k_1 h \alpha^3}\\
\label{E:ConstraintH}
    4 h \frac{\alpha^{\prime}}{\alpha} E_1^{\prime} + \left(2 h^{\prime}  + 8 h \frac{\alpha^{\prime}}{\alpha} \right) E_3^{\prime}
    - 2 \phi^{\prime}\delta\phi^{\prime} h
    +2 \alpha^2 V^{\prime}  \delta\phi & = 0 \\
    E_1^{\prime} + \left(1+\frac{3 k_1^2 v^2}{k^2 h}\right)E_3^{\prime} + E_4^{\prime}
    +\frac{1}{2}\frac{h^{\prime}}{h}E_1
    -3\frac{k_1^2 v^2 h^{\prime}}{k^2 h^2}E_2 \quad \nonumber \\
    {}-\frac{1}{2}\frac{h^{\prime}}{h}\left(1+\frac{3 k_1^2 v^2}{k^2 h}\right)E_3
    +\frac{3}{2}\left(1+\frac{k_1^2 v^2}{k^2 h}
    \right)\phi^{\prime}\delta\phi
    &= -\frac{i k_1 \pi_{\xi}}{k^2 \alpha^3 h}\left(1+ \frac{v^2}{h}\right)
\label{E:Constraint2} \,
\end{align}
\end{subequations}
where we have again written only the leading terms in a $1/k$ expansion. Solving \eqref{E:E2} and \eqref{E:E4} together with the requirement that the metric fluctuations vanish at the asymptotic AdS boundary, we get
\begin{subequations}
\label{Eint}
\begin{align}
  E_2 &= e_2 \int_0^r {d\tilde{r} \over \alpha(\tilde{r})^3} = e_2 \frac{1}{h_1 \alpha_0^3}\left(h-1\right)
\label{Eint2} \\
  E_4 &= e_4 \int_0^r {d\tilde{r} \over h(\tilde{r})
    \alpha(\tilde{r})^3} = e_4 \frac{1}{h_1 \alpha_0^3} \ln h \,, \label{Eint4}
\end{align}
\end{subequations}
where we have used the equation of motion for $h$ \eqref{E:BHeomsa} in the equalities on the right hand side, and defined $h_1 \equiv h^{\prime}(r_H)$. To avoid a logarithmic singularity at the event horizon, one must set $e_4=0$.

The alert reader will have noted that \eno{Eint4} is in precise analogy to \eno{DsolutionsSmallK}.  The horizon boundary condition $e_4=0$ is likewise in analogy to $d_2=0$.  The justification of these boundary conditions is actually somewhat subtle.  At finite $k$, the possible leading near-horizon behaviors of $E_4$ and $D_2$ are $h^{\pm ivk_1/h_1}$.  The minus sign corresponds to an outgoing wave and the plus sign corresponds to an infalling wave.  Usually, the appropriate boundary condition is to suppress the outgoing wave.  But in a small $k$ expansion, $h^{ivk_1/h_1} = 1 + {iv k_1 \over h_1} \log h$.  The relative factor of $k_1$ means that the logarithmic term cannot be visible in a leading small $k$ treatment.  Another way to understand why one must forbid both the infalling and outgoing waves is that the leading order small $k$ analysis does not know about viscosity, so it also doesn't know about the arrow of time; thus the boundary conditions, at this order, should be symmetric between infalling and outgoing waves.

Although it is unimportant for our subsequent analysis, we note for completeness that the $A$ equation of motion completely decouples from the other modes and the scalar fields for all values of $k$.  At order $1/k$, it takes the simple form
 \eqn{Aform}{
    \frac{d}{dr}\left(A^{\prime}h\alpha^3\right) &= 0 \,.
 }
As with $D_2$ and $E_4$, the horizon boundary condition forces $A= \mathcal{O}(1)$.

\subsection{Static peturbations of the horizon}
\label{STATIC}

The equations \eno{E:E2}-\eno{E:ExtraConstraints} are too complicated to solve in general as we did for the $D$ set.  Furthermore, equations \eqref{EEStatic} describe not only the response of the metric to the string, but also static deformations of the black hole background.  Although such deformations are not directly related to the response to the moving quark, it is helpful to describe them explicitly on the dual gravity side as a warm-up to the treatment of sound modes.  More precisely, we want to consider perturbations that are invariant under translations in the ${\bf R}^{3,1}$ directions and rotations in the ${\bf R}^3$ dimensions.  This allows us to set $H_{11} = H_{22} = H_{33}$ and to set all off-diagonal components of the metric to zero.  So $E_2 = E_4 = 0$, while $E_1$ and $E_3$ may be defined as in \eno{E:defE}.  It is straightforward to show that the equations of motion take the form \eno{EEStatic} and that the zero-energy constraint (coming from the $rr$ Einstein equation) takes the form \eno{E:ConstraintH}.  The expectation is that there is only one allowed deformation, and it corresponds to uniformly changing the temperature in the gauge theory.

These equations are still impossible to integrate explicitly for general $V(\phi)$.  However, one may perform a near horizon analysis by switching the independent variable from $r$ to $h$ and expanding
 \eqn{BackgroundExpandStatic}{
  \alpha &= \alpha_0 + \alpha_1 h + \alpha_2 h^2 + \ldots  \cr
  r &= r_H + r_1 h + r_2 h^2 + \ldots  \cr
  \phi &= \phi_0 + \phi_1 h + \phi_2 h^2 + \ldots \,.
 }
We find six linearly independent solutions:
 \eqn{SixSolutionsStatic}{\seqalign{\span\TT\qquad & \span\TL & \span\TR\quad & \span\TL & \span\TR\quad & \span\TL & \span\TR}{
  A: & E_1 &= 1 & E_3 &= 0 & \delta\phi &= 0  \cr
  B: & E_1 &= 0 & E_3 &= 1 & \delta\phi &= 0  \cr
  C: & E_1 &= -{2 \over 3} r_1^2 \alpha_0^2 V'(\phi_0) h &
    E_3 &= -r_1^2 \alpha_0^2 V'(\phi_0) h &
    \delta\phi &= 1 + r_1^2 \alpha_0^2 V''(\phi_0) h  \cr
  D: & E_1 &= -{2 \over 3} r_1^2 \alpha_0^2 V'(\phi_0) h &
    E_3 &= -r_1^2 \alpha_0^2 V'(\phi_0) h &
    \delta\phi &= \left[ 1 + r_1^2 \alpha_0^2 V''(\phi_0) h \right]
       \cr\noalign{\vskip-4\jot}
    &     &\qquad\quad{} \times \log h & &\qquad\quad{} \times \log h &
      &\qquad\quad{} \times \log h  \cr
  E: & E_1 &= {1 \over \sqrt{h}} & E_3 &= 0 & \delta\phi &=
    {2 \over 3} r_1^2 \alpha_0^2 V'(\phi_0) \sqrt{h}  \cr
  F: & E_1 &= 0 & E_3 &= \sqrt{h} & \delta\phi &= -{4 \over 27} r_1^2
    \alpha_0^2 V'(\phi_0) h^{3/2} \,.
 }}
Solutions A and B are exact, while the others, as expressed in \eno{SixSolutionsStatic}, are accurate only to the first two non-trivial orders in small $h$.
Applying the constraint equation \eqref{E:ConstraintH} to the solutions \eno{SixSolutionsStatic}, one finds that the D solution is disallowed, and a relation is enforced between the E and F solutions.  The upshot is that for small $h$,
 \eqn{ConstrainedStatic}{
  E_1 &= {e_1 \over \sqrt{h}} + \tilde{e}_1 - {2 \over 3} r_1^2 \alpha_0^2
    V'(\phi_0) \delta\phi_H h + \ldots  \cr
  E_3 &= e_3 - {2 \over 3} e_1 r_1^2 \alpha_0^2 V(\phi_0) \sqrt{h} -
    \delta\phi_H r_1^2 \alpha_0^2 V'(\phi_0) h + \ldots  \cr
  \delta\phi &= \delta\phi_H + {2 \over 3} e_1 r_1^2 \alpha_0^2
    V'(\phi_0) \sqrt{h} + \delta\phi_H r_1^2 \alpha_0^2 V''(\phi_0)
      h + \ldots \,,
 }
where $e_1$, $\tilde{e}_1$, $e_3$, and $\delta\phi_H$ are integration constants.  In principle, three of these constants can be fixed in terms of the fourth by integrating $E_1$, $E_3$, and $\delta\phi$ out to the conformal boundary of $AdS_5$ and imposing boundary conditions there corresponding to the absence of any deformations of the lagrangian.  This means that the boundary conditions on the asymptotic AdS boundary are such that $E_1$ and $E_3$ are $\mathcal{O}(r^4)$ near the conformal boundary, while $\delta\phi$ is $\mathcal{O}(r^\Delta)$, where $\Delta$ is the dimension of the operator dual to $\phi$.\footnote{There is a subtlety when $\Delta$ is between $2$ and $3$: different QFT's exist, which are conformal at least in a large~$N$ limit, which allow $\Delta$ to be exchanged with $4-\Delta$ \cite{Klebanov:1999tb}.  The simplest assumption is $\Delta \geq 2$, but the precise form of the boundary condition on $\delta\phi$ does not matter in our analysis.}  Once these conditions from the conformal boundary are imposed, what remains is just one static deformation of the black hole, as expected on general grounds.\footnote{Intriguingly, we do not need to impose any boundary conditions at the horizon as we have in the discussion following \eqref{Eint}. It seems that at order $1/k$ the constraint equation \eqref{E:ConstraintH} is enough to ensure that $\delta \phi$ and all the $E_i$'s are finite at the horizon.}

For the $AdS_5$-Schwarzschild background, the differential equations \eno{EEStatic} can be integrated out to the conformal boundary of anti-de Sitter space, and the boundary conditions there can be imposed explicitly.  They are $e_1 = -\tilde{e}_1 = -2e_3$, with no constraint on $\delta\phi_H$ because the scalar and graviton perturbations decouple.\footnote{The $AdS_5$-Schwarzschild case is special in that a constant shift of the dilaton is an exact deformation.  So only two additional constraints arise from the conformal boundary.  Usually there are three.}  To show all this, simply note that the terms shown explicitly in \eno{ConstrainedStatic} provide an exact solution for AdS${}_5$-Schwarzschild.  For other backgrounds, one usually cannot explicitly perform the integration from the horizon to the conformal boundary.  Generically, the scalar and graviton mix, and the three constraints at the conformal boundary involve all of $e_1$, $\tilde{e}_1$, $e_3$, and $\delta\phi_H$.

Since we can not integrate the equations, we proceed by expressing the various integration constants in terms of the entropy and temperature.
The solutions \eqref{ConstrainedStatic} imply that the metric fluctuations are given by
 \eqn{MetricPertStatic}{
  H_{00} =
    -{2 \over 3} e_1 \sqrt{h} + {2 \over 3} (e_3-\tilde{e}_1) h +
    \mathcal{O}(h^{3/2}) \qquad
   H_{11} = {2 \over 3} e_3 + \mathcal{O}(h^{1/2}) \,
 }
(recall that $H_{11} = H_{22} = H_{33}$.)  Note that despite the $1/\sqrt{h}$ singular terms in \eqref{ConstrainedStatic}, the fluctuations \eno{MetricPertStatic} are finite at the horizon.  In \eno{MetricPertStatic} we have suppressed all terms beyond the ones that affect the variation in the entropy and temperature.  The entropy is straightforward to compute: the perturbed metric to order $\kappa_5^2$ takes the form
 \eqn{PerturbedMetStatic}{
  ds^2 = \alpha^2 \left[ -(h-\kappa_5^2 H_{00}) dt^2 +
    (1+\kappa_5^2 H_{11}) d\vec{x}^2 +
    {dr^2 \over h} \right] \,.
 }
So the entropy per coordinate volume of $\vec{x}$ is
 \eqn{EntropyDensity}{
  s = {2\pi \alpha_0^3 \over \kappa_5^2}
    (1 + \kappa_5^2 H_{11})^{3/2} \,
 }
evaluated at the horizon.  Thus
 \eqn{deltaSoverS}{
  {\delta s \over s} =
    \kappa_5^2 \lim_{h\to 0} {3 \over 2} H_{11} =
    \kappa_5^2 e_3 \,.
 }
Determining the variation of the temperature is a bit more subtle because $G_{tt}$ and $G_{rr}$ depend non-analytically on $h$ after the perturbation.  One way to proceed is to change variables from $r$ to $\tilde{r}$ in such a way that the perturbed metric becomes
 \eqn{ChangedGauge}{
  ds^2 = \tilde\alpha^2 \left[ -\tilde{h} dt^2 + d\vec{x}^2 +
    {d\tilde{r}^2 \over \tilde{h}} \right] \,.
 }
One easily finds
 \eqn{FoundTildeVars}{
  \tilde\alpha &= \alpha \left( 1 +
    {\kappa_5^2 \over 2} H_{11} \right)  \cr
  \tilde{h} &= h - \kappa_5^2 H_{00} - \kappa_5^2 h H_{11}
     \cr
  {d\tilde{r} \over dr} &= 1 - {\kappa_5^2 \over 2}
    {H_{00} \over h} -
    \kappa_5^2 H_{11} \,,
 }
all up to $\mathcal{O}(\kappa_5^4)$ corrections which we ignore.  The Hawking temperature of the unperturbed background \eno{BHansatz} is easily seen to be $1/4\pi r_1$.  Suppose we expand $\tilde{r}$ in a power series in $\tilde{h}$:
 \eqn{rTildeExpand}{
  \tilde{r} = \tilde{r}_H + \tilde{r}_1 \tilde{h} +
    \tilde{r}_2 \tilde{h}^2 + \ldots \,.
 }
Then the Hawking temperature of the perturbed background is
 \eqn{PerturbedHawking}{
  T = -{1 \over 4\pi \tilde{r}_1} \,.
 }
Therefore
 \eqn{deltaToverT}{
  {\delta T \over T} = 1-{\tilde{r}_1 \over r_1} =
    \kappa_5^2
     \lim_{h\to 0} \left[ -{dH_{00} \over dh} + {H_{00} \over 2h} -
      h {dH_{11} \over dh} \right] =
    \kappa_5^2 {\tilde{e}_1-e_3 \over 3} \,.
 }
According to simple thermodynamic arguments summarized in \cite{Buchel:2005nt}, as long as there are no chemical potentials present, the speed of sound should be
 \eqn{SpeedOfSound}{
  c_s^2 = {d\log T \over d\log s} = {1 \over 3} \left(
    {\tilde{e}_1 \over e_3} - 1 \right) \,.
 }
As a check, one may subsitute $\tilde{e}_1=2e_3$ for AdS-Schwarzschild and recover $c_s^2 = 1/3$.  It will also prove interesting to note that the first law of thermodynamics predicts
 \eqn{dEfirstLaw}{
  \delta {\cal E} = T\delta s = -{3\alpha_0^3 \over 4r_1} H_{11}
    = -{\alpha_0^3 \over 2r_1} e_3 \,,
 }
where ${\cal E}$ is the energy density and we again assume that no chemical potential is present.

\subsection{Small momentum asymptotics}
\label{SONIC}

A striking feature of the full $\mathcal{O}(1/k)$ equations \eno{E:E2}-\eno{E:ExtraConstraints} is that the equations of motion for $E_1$, $E_3$, and $\delta\phi$, together with the zero energy constraint \eno{E:ConstraintH}, are the same as for static deformations. The two extra non homogeneous constraints \eqref{E:Constraint1} and \eqref{E:Constraint2} will determine $e_2$ and one other integration constant. Heuristically, this means that we can describe sound waves as a modulation of the static deformation, plus some profile for $E_2$ that incorporates non-zero Poynting vector.  Technically, it means we can set
 \eqn{SoSagain}{
  \tilde{e}_1 = (1 + 3 c_s^2) e_3 \,.
 }
This is not really a constraint on $\tilde{e}_1$, instead we are simply trading $\tilde{e}_1/e_3$ for $c_s^2 \equiv d\log T / d\log s$ as a parameter for describing temperature-deformations. Only by fully integrating the differential equations \eno{EEStatic} and~\eno{ConstrainedStatic} and imposing appropriate constraints at the conformal boundary could we determine $c_s^2$.  Plugging \eno{ConstrainedStatic} and~\eno{Eint} into \eno{E:ExtraConstraints}, and imposing $e_4=0$ as well as \eno{SoSagain}, one obtains a pair of linear equations in $e_2$ and $e_3$.  Solving these equations leads to
 \eqn{FoundEtwo}{
  e_2 = {\pi_\xi \over ik_1} {k_1^2 + c_s^2 k^2 \over
    v^2 k_1^2 - c_s^2 k^2}  \qquad
  e_3 = {2 r_1 k_1 \pi_\xi \over i\alpha_0^3} {1+v^2 \over
    v^2 k_1^2 - c_s^2 k^2} \,.
 }

Using \eno{GotDone} and~\eno{FoundEtwo} in \eqref{HolographicPoynting} we obtain the Poynting vector to leading order in small $k$:
 \eqn{FinalPoynting}{
  S_1 &= -{\pi_\xi \over i v k_1} +
    {i c_s^2 k_1 \pi_\xi \over v} {1+v^2 \over
     v^2 k_1^2 - c_s^2 k^2} + \mathcal{O}(1) \cr
  S_\perp &= {i c_s^2 k_\perp \pi_\xi \over v} {1+v^2 \over
     v^2 k_1^2 - c_s^2 k^2} + \mathcal{O}(1) \,.
 }
When $c_s = 1/\sqrt{3}$, \eno{FinalPoynting} reduces to the leading order results found in \cite{Gubser:2007ga}.
It is natural to identify the contribution of sound modes as
 \eqn{SsDef}{
  \vec{S}^{\rm s} = \begin{pmatrix} S_1^{\rm s} &
    S_\perp^{\rm s} & 0 \end{pmatrix} \,,
 }
where
 \eqn{SoundModes}{
  S^{\rm s}_1 = {k_1 \over k_\perp} S^{\rm s}_\perp =
    {i c_s^2 k_1 \pi_\xi \over v} {1+v^2 \over
     v^2 k_1^2 - c_s^2 k^2} \,.
 }
Defining $\vec{S}^{\rm d}$ as in \eno{SimpleWake}, we see that $\vec{S} = \vec{S}^{\rm s} + \vec{S}^{\rm d}$.

In order to completely characterize the energy dissipation, we need also to compute the fluctuations in the energy density.
This is difficult to work out in a general setting because the Brown-York tensor probably does not correctly capture $\langle T_{00} \rangle$ owing to effects from holographic renormalization. Also one apparently has to solve for $E_1$ and $E_3$ out to the boundary, which seems impossible.  But there is a simple shortcut---though admittedly heuristic: it is simply to use the thermodynamic relation $\delta {\cal E} = T\delta s$ in the $\vec{k}$-th Fourier mode.  This should work at leading order in small $k$.  The result, carried over directly from \eno{dEfirstLaw}, is
 \eqn{ThermoDeltaE}{
  \delta {\cal E} = -{\alpha_0^3 \over 2r_1} e_3
    = i k_1 \pi_\xi {1+v^2 \over v^2 k_1^2 - c_s^2 k^2} \,.
 }
When $c_s = 1/\sqrt{3}$, \eno{ThermoDeltaE} reduces to the leading order results found in \cite{Gubser:2007ga}.

\subsection{Non-conservation of the stress tensor}
\label{NONCONSERVATION}

The results \eno{FinalPoynting} and~\eno{ThermoDeltaE} are enough information to determine the first row of the stress tensor completely at small $k$.  More precisely, consider the subtracted stress energy tensor expectation value at time $t=0$ (still in the rest frame of the plasma):
 \eqn{deltaT}{
  \delta T^{mn} \equiv \langle T^{mn}(0,\vec{x}) \rangle -
    \langle T^{mn} \rangle_{\rm bath} \,,
 }
where $\langle T^{mn} \rangle_{\rm bath}$ is from the infinite, static, thermal medium.  What we have computed, in Fourier space, with $k_3=0$ and $k_2=k_\perp>0$, is
 \eqn{deltaTfromES}{
  \delta T^{0m} = \begin{pmatrix} \delta {\cal E} & S_1 & S_\perp &
    0 \end{pmatrix} \,.
 }
(Recall that we use $\vec{S}$ for the response of the Poynting vector to the string instead of $\delta\vec{S}$.)
Rotational invariance around the $\hat{1}$ axis determines $\delta T^{0m}$ for other values of $\vec{k}$.  As discussed in \cite{Gubser:2007ga,Friess:2006fk}, the failure of the conservation law $\partial_m \delta T^{mn} = 0$ is a good measure of energy dissipation from the quark.  The non-conservation law can be expressed in Fourier space as
 \eqn{Nonconservation}{
  i k_m \delta T^{0m} = P^{\rm t} \,,
 }
where
 \eqn{kmDef}{
  k_m = \begin{pmatrix} -vk_1 & k_1 & k_\perp & 0 \end{pmatrix} \,
 }
(setting $k_0=-vk_1$ comes from the co-moving ansatz.)  The right hand side of \eno{Nonconservation} was computed in \eqref{TotalEnergyLoss}. It is the total rate at which the quark deposits energy into the plasma: a positive quantity.  A check on our computations so far is to plug \eno{FinalPoynting}, \eno{ThermoDeltaE}, and \eno{deltaTfromES} into \eno{Nonconservation} and check that the right hand side agrees with \eno{TotalEnergyLoss}.  It does.

As in \cite{Gubser:2007ga}, we associate $\delta {\cal E}$ entirely with dissipation into sound waves.  This makes sense because, in a linearized hydrodynamical analysis, the diffusion wake shows up only in the Poynting vector.  Thus we have a splitting
 \eqn{deltaTsplits}{
  \delta T^{0m} = \delta T^{0m}_{\rm sound} +
    \delta T^{0m}_{\rm diffuse} \,,
 }
where
 \eqn{deltaTsplitting}{
  \delta T^{0m}_{\rm sound} &= \begin{pmatrix} \delta {\cal E} &
    S_1^{\rm s} & S_\perp^{\rm s} & 0 \end{pmatrix}
     \cr\noalign{\vskip2\jot}
  \delta T^{0m}_{\rm diffuse} &= \begin{pmatrix} 0 &
    S_1^{\rm d} & 0 & 0 \end{pmatrix} \,.
 }
There is an associated splitting of the non-conservation effect:
 \eqn{NonConSplitting}{
  i k_m T^{0m}_{\rm sound} &= P^{\rm s}
    \cr\noalign{\vskip2\jot}
  i k_m T^{0m}_{\rm diffuse} &= P^{\rm d} \,.
 }
We regard \eno{NonConSplitting} as a definition of $P^{\rm s}$ and $P^{\rm d}$.  It is immediate from the preceding discussion that
 \eqn{RatioOneMoreTime}{
  {P^{\rm s}} : {P^{\rm d}} :
    P^{\rm t} = 1+v^2 : -1 : v^2 \,,
 }
which is the result \eno{EnergyRatio} that we advertised from the start.

It is clear that in splitting up $\delta T^{0m}$ as in \eno{deltaTsplitting} we are making choices rather than doing something inevitable.
To justify this splitting we note that in linearized hydrodynamics, neglecting viscosity, the Fourier space form describing sound waves involves $\omega^2 - c_s^2 k^2$, while the analogous form for the diffusion wake is just $\omega$ (see for example \cite{Casalderrey-Solana:2004qm}). This precisely coincides with our choice of sound and diffusion modes in equations \eqref{SimpleWake}, \eqref{SoundModes} and \eqref{ThermoDeltaE}.  The diffusion part of the stress tensor is exactly the long, narrow stream of plasma behind the quark, which we justified in a different way in the discussion leading up to \eno{SimpleWake}.

Furthermore, when $v > c_s$, the split we use leads to exponential suppression of all components of $\delta T^{0m}_{\rm sound}$ and $\delta T^{0m}_{\rm diffuse}$ for $x^1 > 0$, which is sensible because this region is causally inaccessible to sound waves emanating from the quark.  Another natural split is to associate contributions to the stress tensor from the $D$ set purely with the diffusion wake, and contributions from the $E$ set purely with sound modes.  This is the same transverse/longitudinal split discussed in \cite{Casalderrey-Solana:2004qm} and it leads to power law tails of components of the stress tensor cancelling delicately between diffusion and sound modes.

It is interesting to note that when $c_s \to 0$, the total perturbation in the stress tensor becomes
 \eqn{ZeroSpeed}{
  \delta T^{0m}(\vec{k}) = {i\pi_\xi \over vk_1} \begin{pmatrix}
    {1+v^2 \over v} & 1 & 0 & 0 \end{pmatrix} \,,
 }
with our usual convention that $k_3=0$ and $k_2=k_\perp>0$.  In position space,
 \eqn{ZeroSpeedPosition}{
  \delta T^{0m}(0,\vec{x}) = {\pi_\xi \over v} \theta(-x^1)
    \delta(x^2) \delta(x^3) \begin{pmatrix}
     {1+v^2 \over v} & 1 & 0 & 0 \end{pmatrix} \,,
 }
which describes energy deposition and flow only in the path of the quark.  There is of course some viscosity broadening, presumably visible at higher orders in $k$, but no pressure broadening---because there is no pressure.  Sound waves don't really exist in this limit.  But the diffusion wake, as defined in \eno{SimpleWake}, is completely insensitive to the speed of sound.  Its form and its strength appear to be a universal feature of string theory constructions starting from Einstein gravity coupled to strings and scalars.

\section{Relation to linearized hydrodynamics}
\label{HYDRO}

The stress energy tensor of the thermal plasma is conserved except at the location of the quark:
 \eqn{ConservationPosition}{
  \partial_m \delta T^{mn} = J^n \delta(x^1-vt)\delta(x^2)\delta(x^3) \,.
 }
In Fourier space,
\begin{equation}
\label{E:ConservationExact}
    i k_{m} \delta T^{mn} = J^n \,,
\end{equation}
where
 \eqn{kmAgain}{
  k_m = \begin{pmatrix} -vk_1 & k_1 & k_\perp & 0 \end{pmatrix}
 }
as in \eno{kmDef}.  The $n=0$ component of \eno{E:ConservationExact} is precisely \eno{Nonconservation}.  From \eqref{DragForce} and \eqref{TotalEnergyLoss} we read off
 \eqn{Jresult}{
  J^n = \begin{pmatrix} v \pi_{\xi} & \pi_{\xi} & 0 & 0 \end{pmatrix}
   \,.
 }
We emphasize that our derivation of \eno{E:ConservationExact}, for $n=0$, did not directly use a hydrodynamic approximation: it came from the small $k$ limit of an AdS/CFT computation.  The purpose of this section is to understand how the AdS/CFT computation matches with linearized hydro, including (for ${\cal N}=4$) a sub-leading effect in small $k$.

The constitutive relations for linearized hydro amount to
\begin{equation}
\label{E:HydroApproximation}
    \delta T_{{\rm hydro},{ij}}  = c_s^2 \epsilon \delta_{ij} - \frac{3}{4}\Gamma_s \left(i k_i S_j + i k_j S_i - \frac{2}{3}\delta_{ij} i k_l S^l\right) \,,
\end{equation}
where
 \eqn{SiEpsilonHydro}{
  \epsilon = \delta T_{\rm hydro}^{00} \qquad S_i = \delta T_{\rm hydro}^{i0} \,.
 }
(Recall that we use mostly plus signature.)  The equations of motion for linearized hydrodynamics are
\begin{equation}
\label{E:ConservationHydro}
    i k_{m} \delta T^{mn}_{\rm hydro} = J^{n}_{\rm hydro} \,.
\end{equation}
But now there is no reason why the source term $J^n_{\rm hydro}$ should be pointlike.  Instead, in position space, $J^n_{\rm hydro}$ probably has width comparable to $\Gamma_s$ (the attenuation length) and must be parametrized in some fashion.  The authors of \cite{Casalderrey-Solana:2004qm,CasalderreySolana:2006sq} used an equivalent of
 \eqn{JhydroParametrize}{
  J_{\rm hydro}^n = \begin{pmatrix} e_0 & g_0 + k_1 g_1
   & k_\perp g_1 & 0 \end{pmatrix} \,.
 }
In \cite{Casalderrey-Solana:2004qm,CasalderreySolana:2006sq} two scenarios where considered. In scenario~1, $e_0 = v g_0$ is a Gaussian in real space with width $\Gamma_s$, while $g_1=0$. Scenario 2 corresponds to the opposite extreme, $e_0=g_0=0$ and $g_1$ a Gaussian with width $\Gamma_s$.

A natural though non-unique prescription in matching our results to linearized hydro is to set
 \eqn{FirstRowMatches}{
  \delta T_{\rm hydro}^{m0} = \delta T^{m0}
 }
for $m=0,1,2,3$ and then to use \eno{E:HydroApproximation} and~\eno{SiEpsilonHydro} to determine the other components of $\delta T_{\rm hydro}^{mn}$.\footnote{One possible alternative prescription for determining $\delta T_{\rm hydro}^{m0}$ is to require that at order $\mathcal{O}(k)$, $J_{\rm hydro}^i$ has only longitudinal components. We've checked that the results we describe below remain unchanged under such a redefinition.}  If we follow this procedure to order $\mathcal{O}(k^{-1})$, see \eno{FinalPoynting} and \eno{ThermoDeltaE}, we find
\begin{equation}
\label{E:Thydro0}
    \delta T^{mn}_{\rm hydro}
    =
    \begin{pmatrix}
    \frac{i k_1 \left( 1 + v^2 \right)}{-c_s^2 k^2  +k_1^2 v^2}
        & -\frac{1}{i k_1 v} - \frac{i c_s^2 k_1 (1+v^2)}{v(c_s^2 k^2 - k_1^2 v^2)}
        & -\frac{i c_s^2 k_{\bot} (1+v^2)}{v(c_s^2 k^2 - k_1^2 v^2)}
        & 0 \\
        -\frac{1}{i k_1 v} - \frac{i c_s^2 k_1 (1+v^2)}{v(c_s^2 k^2 - k_1^2 v^2)}
        & \frac{i k_1 c_s^2\left( 1 + v^2 \right)}{-c_s^2 k^2  +k_1^2 v^2}
        & 0 & 0 \\
        -\frac{i c_s^2 k_{\bot} (1+v^2)}{v(c_s^2 k^2 - k_1^2 v^2)} &
        0 &
        \frac{i k_1 c_s^2\left( 1 + v^2 \right)}{-c_s^2 k^2  +k_1^2 v^2} & 0 \\
        0 & 0 & 0 &
        \frac{i k_1 c_s^2\left( 1 + v^2 \right)}{-c_s^2 k^2  +k_1^2 v^2} \\
    \end{pmatrix}
    \pi_{\xi}
    +\mathcal{O}(1)\,.
\end{equation}
If we plug \eno{E:Thydro0} into \eno{E:ConservationHydro}, we obtain an expression for $J_{\rm hydro}^n$ coinciding with \eno{JhydroParametrize} with
 \eqn{Pointlike}{
  e_0 = v\pi_\xi + \mathcal{O}(k) \qquad
  g_0 = \pi_\xi + \mathcal{O}(k) \qquad
  g_1 = \mathcal{O}(1) \,.
 }
Fourier transforming the leading terms shown in \eno{Pointlike} leads to position results which are delta-functions, as in \eno{ConservationPosition}.  Because of the explicit factors of $k$ in \eno{JhydroParametrize}, non-vanishing $e_0$ and $g_0$ tend to dominate over non-vanishing $g_1$ at sufficiently long distances.  More specifically, given $e_0$, $g_0$ and $g_1$ in Fourier space, one may define a dimensionless figure of merit,
 \eqn{gammaDef}{
  \gamma_1 = {1 \over R} \left| {g_1(k=1/R) \over e_0(k=1/R)} \right| \,,
 }
such that $\gamma_1 \ll 1$ favors scenario 1 and $\gamma_1 \gg 1$ favors scenario 2.  $R$ is a typical distance at which one is interested in the strength of the flow induced by $e_0$ and $g_0$ as compared to the flow induced by $g_1$.\footnote{J.~Casalderrey-Solana has suggested to us that $\gamma_1$ could be changed by sending $R \to \sqrt{\Gamma_s R}$ in \eno{gammaDef}.  This results in a rough expectation $\gamma_1 \sim \sqrt{\Gamma_s/R}$ in place of \eno{gammaExpect}.}  (We assume that $e_0$ and $g_0$ have similar magnitudes.)  If $J^n_{\rm hydro}$ is localized at a scale $\Gamma_s$, then for a source such as the one we have described, where $e_0$ and $g_0$ are non-vanishing at leading order, one expects
 \eqn{gammaExpect}{
  \gamma_1 \sim {\Gamma_s \over R} \,.
 }
Because one only trusts linearized hydro for $R \gg \Gamma_s$, our results correspond to scenario~1.  To get scenario~2, one needs $\gamma_1 \gg 1$.

The conclusion \eno{gammaExpect} was based on dimensional reasoning and so should be treated with some caution.  We can calculate a precise value for $\gamma_1$, as well as more information about the structure of the source terms $e_0$ and $g_0$, by calculating the stress tensor at the first subleading order in $k$.  That is, we expand
 \eqn{ExpandTmn}{
  \delta T^{mn} &= {\delta T^{(0)\,mn} \over k} + \delta T^{(1)\,mn} + \mathcal{O}(k)  \cr
  \delta T_{\rm hydro}^{mn} &= {\delta T^{(0)\,mn}_{{\rm hydro}} \over k} +
    \delta T^{(1)\,mn}_{{\rm hydro}} + \mathcal{O}(k)
 }
and impose
\eno{E:HydroApproximation} and \eno{FirstRowMatches} at order $\mathcal{O}(1)$ to get $\delta T^{(1)\,mn}_{{\rm hydro}}$. These subleading contributions to the energy momentum tensor are currently available only for the $\mathcal{N}=4$ theory \cite{Friess:2006fk}. Adapting the notation in \cite{Friess:2006fk} to the current conventions, we find that
 \eqn{ToneCompare}{
    \delta T^{(1)\, mn}= \delta T_{\rm hydro}^{(1)\, mn}  + \hbox{diagonal}\begin{pmatrix}0 & -2 & 1 & 1\end{pmatrix}\frac{v \pi_{\xi}}{3\pi T}.
 }
This corresponds to
\begin{equation}
\label{E:Jversion1}
    J^n = \begin{pmatrix} v & 1 - 2v k_1 \Gamma_s  & v k_{\bot} \Gamma_s & 0 \end{pmatrix}\pi_{\xi}+\mathcal{O}(k^2) \,,
\end{equation}
where we have used $\Gamma_s = 1/3\pi T$.  From \eno{E:Jversion1} we read off
 \eqn{SourceValues}{
  e_0(\vec{k}) = \pi_\xi v + \mathcal{O}(k^2) \qquad
   g_0(\vec{k}) = \pi_\xi (1 - 3 v k_1 \Gamma_s) + \mathcal{O}(k^2) \qquad
   g_1(\vec{k}) = \pi_\xi v \Gamma_s + \mathcal{O}(k) \,.
 }
These results agree with the expectation that $J^n_{\rm hydro}$ has structure on the scale $\Gamma_s$, and it implies that
 \eqn{GotGamma}{
  \gamma_1 = {\Gamma_s \over R} + \mathcal{O}(R^{-2}) \,.
 }
Taking $\Gamma_s = 0.07\,{\rm fm}$ and $R = 7\,{\rm fm}$ to make a rough comparison with heavy ion collisions, one finds $\gamma_1 = 0.01$.


It seems likely that $\gamma_1$ will be be small for phenomenologically interesting values of $R$ even for AdS/CFT constructions including scalars, unless perhaps the scalar lagrangian has some very special or extreme form.  Nevertheless, it would be interesting to see what form these subleading expressions take in, for example, the $\mathcal{N}=2^{\star}$ theory \cite{Pilch:2000ue,Buchel:2003ah,Buchel:2007vy}, or the casacading gauge theory \cite{Klebanov:2000hb,Gubser:2001ri,Mahato:2007zm,Aharony:2007vg}.  It seems to us more likely that $\gamma_1$ could be altered significantly by considering a different source from the heavy quark we have employed throughout.

\section{Jet-splitting and the challenge of diffusion wakes}
\label{DISCUSSION}

In this section we attempt to bridge between our results from AdS/CFT and the phenomenological literature on heavy ion collisions.  In doing so we should bear in mind all the usual caveats to such comparisons.  In particular: the AdS/CFT results apply in the limit of large $N$ and large 't~Hooft coupling $g_{YM}^2 N$; the quark in our treatment is infinitely heavy and has constant velocity; and the holographic dual to QCD is not known.

\begin{figure}
  \centerline{\includegraphics[width=7in]{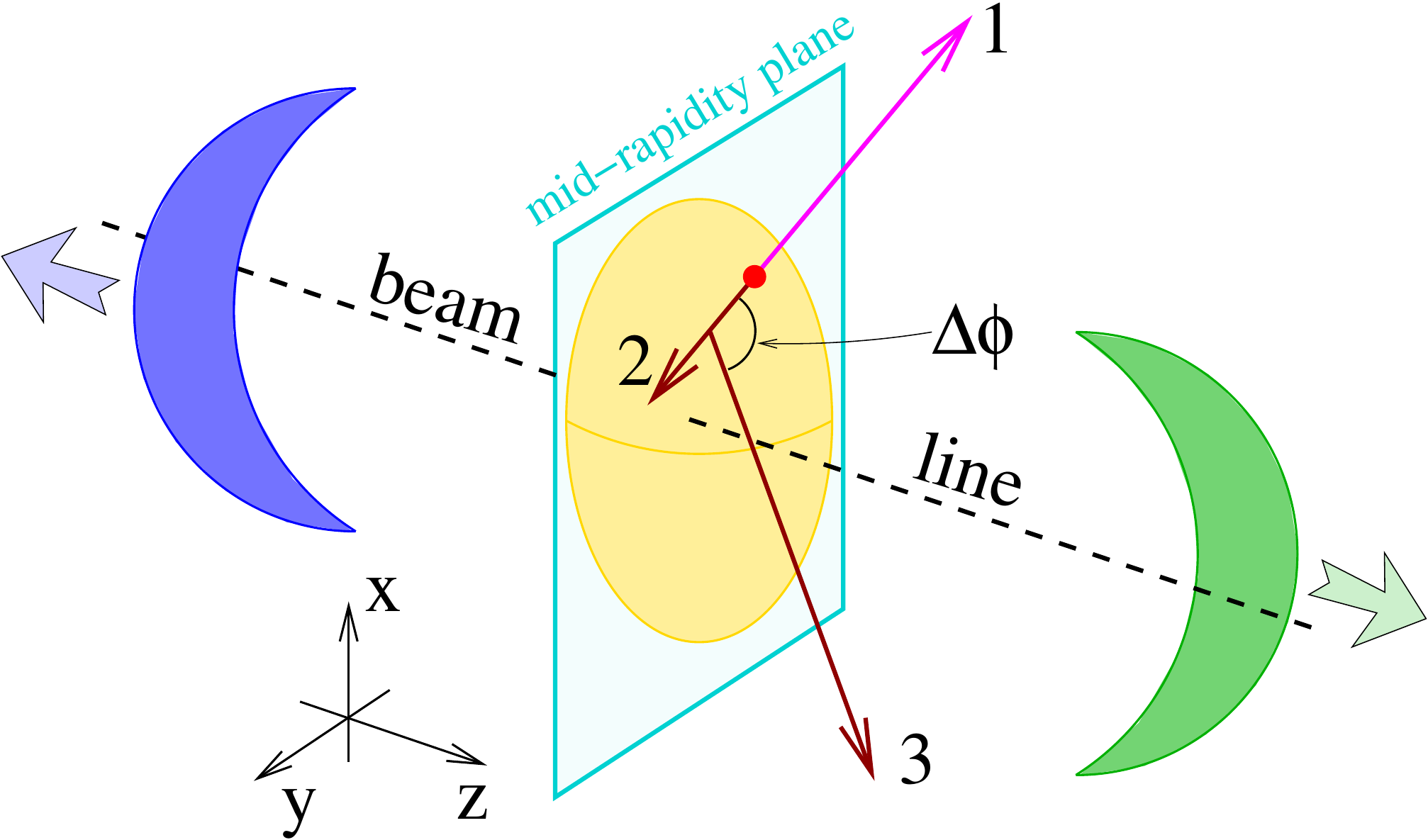}}
  \caption{A cartoon of a hard process in a heavy ion collision.  The blue and green crescents are the parts of the nuclei that didn't interact.  The gold region is the thermalizing medium.  The red dot is the vertex of a hard process that occurs early in the collision, producing two partons with back-to-back momenta in the azimuthal direction.  For simplicity we show all particles in the $x$-$y$ plane meaning at mid-rapidity.  If particles $1$ and $2$ escape to form hadrons, it would correspond to $\Delta\phi \approx \pi$.  If instead particle $2$ stops in the medium but emits a high-angle secondary which is then observed as $3$, then $\Delta\phi$ is significantly different from $\pi$.}\label{azimuth}
 \end{figure}

In heavy ion collisions, a well studied experimental probe of the interaction of a hard parton with the medium is the relative azimuthal angle $\Delta\phi$ between two energetic hadrons emitted from the interaction region.  Histograms of hadron pairs always show a peak at $\Delta\phi=0$, most likely meaning that the two hadrons were part of the same jet.  In proton-proton collisions and sufficiently peripheral heavy ion collisions where the extent of the medium is small, another peak at $\Delta\phi = \pi$ arises, owing presumably to events where two hard partons are created with back-to-back momenta---at least, back-to-back in the azimuthal direction.  See figure~\ref{azimuth}.

Jet-broadening or jet-splitting refers to the change in shape of the peak at $\Delta\phi=\pi$.
Di-hadron histograms published by the PHENIX collaboration in \cite{Adler:2005ee} show a definite double-peaked structure: a minimum at $\Delta\phi = \pi$ and peaks at $\Delta\phi \approx \pi \pm 1.2$.
On the other hand, the di-hadron histograms published by the STAR collaboration in \cite{Adams:2005ph} do not exhibit a pronounced double-peaked structure and are discussed in that paper in terms of softening in $p_T$, broadening in $\Delta\phi$ and rapidity, and thermalization.   Momentum cuts were significantly different in these two studies:
 \eqn{MomentumCuts}{\seqalign{\span\TT: \qquad & \span\TR}{
  STAR & 0.15\,{\rm GeV} < p_T^{\rm assoc} <
          4\,{\rm GeV} < p_T^{\rm trig} < 6\,{\rm GeV}  \cr
  PHENIX & 1\,{\rm GeV} < p_T^{\rm assoc} < 2.5\,{\rm GeV} <
          p_T^{\rm trig} < 4\,{\rm GeV} \,.
 }}
STAR's pseudo-rapidity acceptance, $|\eta|<1$, is also significantly greater than PHENIX's, $|\eta| < 0.35$.  Subsequent experimental analyses, including \cite{Ulery:2005cc,Adare:2006nr,Horner:2007gt,Ulery:2007zb,Jia:2007sf}, support the conclusion that at least for intermediate momentum hadrons, similar to the PHENIX cuts shown in \eno{MomentumCuts}, splitting of the away side peak does occur.

A candidate explanation for jet-splitting is that a hard parton traveling through a quark-gluon plasma (QGP) loses energy in large part by producing a sonic boom \cite{Casalderrey-Solana:2004qm,Stoecker:2004qu}.  The sonic boom would then propagate through the plasma until freeze-out converts it into an excess of hadrons emitted in roughly the direction of the Mach angle, relative to the direction of the original parton.
There are theoretical alternatives (see for example \cite{Ruppert:2005uz,Koch:2005sx,Polosa:2006hb,Vitev:2005yg}), but our string theory results bear less directly upon them. In \cite{Casalderrey-Solana:2004qm}, two scenarios were presented for the distribution of energy into sound modes and diffusion modes. In section~\ref{HYDRO} we have shown that in scenario~1 the relative strength of the sonic boom and the diffusion wake is the same as in our string theory analysis. After Cooper-Frye hadronization on a surface of fixed time, it shows no jet-splitting due to the relatively strong diffusion wake. In scenario~2, only longitudinal modes are sourced by the energy momentum tensor so the diffusion wake is entirely suppressed and a jet-splitting effect is recovered after Cooper-Frye hadronization.  Further work \cite{CasalderreySolana:2006sq} showed that for scenario~2 to match to data, the energy loss per distance needs to be chosen quite large---perhaps unrealistically so.  Another study \cite{Chaudhuri:2005vc} included expansion of the plasma and used source terms similar to scenario~1.  It also concluded that a minimum at $\delta\phi = \pi$ is hard to achieve.  In \cite{Renk:2005si,Renk:2006mv}, a model including Mach cone propagation through the QGP has been argued to fit experimental data provided the fraction of energy going into the sonic boom is $0.75$ or more.  But it is not yet clear to us how this fraction is related to the relative strength of the sonic boom and the diffusion wake as we compute it.

The robustness of the diffusion wake in the string theory construction we have explored favors scenario~1.  It is therefore a concern that this scenario seems less likely to agree with data.  It is natural for us to hope that a diffusion wake really is generated when a hard parton traverses a quark-gluon plasma, but that it is erased or disguised in the data.  More realistic theoretical treatments should include expansion of the plasma, as in \cite{Chaudhuri:2005vc,Renk:2005si,Renk:2006mv} and perhaps also a more sophisticated treatment of hadronization; as compared to the sonic boom, the diffusion wake is deeper inside the expanding medium, so it has more chance to broaden, soften, and thermalize before freeze-out.  Some hints of this were observed in \cite{Chaudhuri:2005vc}.  There also issues of detector acceptance which may be relevant, see for example the discussion in \cite{Renk:2006mv}. Finally,  we should keep firmly in mind our starting assumption that the quark is very massive.  If the trigger particle could be tagged as originating from a heavy quark, then the distribution of associated hadrons would be more directly related to our analysis.  Studies cited thus far do not include flavor tagging; see however \cite{Lin:2007rn}.

In summary: We have argued that the strength of the diffusion wake relative to the drag force is universal in AdS/CFT constructions based on a trailing string in a background supported by arbitrary scalar matter.  This universality is in some ways comparable to the universality of the $\eta/s$ ratio \cite{Buchel:2003tz,Kovtun:2004de,Buchel:2004di} (see also \cite{Buchel:2007mf}): it arises because of a decoupling of certain metric perturbations from the scalar dynamics, and it may suffer corrections in inverse powers of $N$ and $g_{YM}^2 N$.  Also, it is worth remembering that our slightly weaker result on the universality of dissipation through sound modes does not specify how well this dissipation is focused at the Mach angle.  Phenomenologically, it may be challenging to accommodate as strong a diffusion wake as we predict.

\section*{Acknowledgments}

The work of S.~Gubser was supported in part by the Department of Energy under Grant No.\ DE-FG02-91ER40671, and by the Sloan Foundation.  A.~Yarom is supported in part by the German Science Foundation and by the Minerva foundation. S.S.G.~thanks T.~Hemmick, S.~Pratt, and B.~Jacak for useful discussions. A.Y.~thanks J.~Casalderrey-Solana, M.~Haack, E.~Shuryak and D.~Teaney for useful discussions. A.~Y. is grateful to the 5th Simons workshop at YITP and the Princeton University Physics Department for hospitality.

\begin{appendix}
\section{Holographic renormalization}
\label{HOLORG}

The prescription for calculating the expectation value of a gauge theory operator $\langle \mathcal{O}_{\Phi} \rangle$ dual to a field $\Phi$ in the supergravity limit of string theory is to identify the generating function of $\langle \mathcal{O}_{\Phi} \rangle$ with the supergravity partition function \cite{Gubser:1998bc,Witten:1998qj}. Sources for the operator $\langle \mathcal{O}_{\Phi} \rangle$ in the gauge theory are identified with boundary values of $\Phi$ (modulo some issues with low dimension operators \cite{Klebanov:1999tb}.) In the saddle point approximation, calculating one point functions reduces to varying the supergravity action with respect to the boundary values of the fields $\Phi$ near the AdS boundary.
\begin{equation}
\label{E:OPhiVEV}
        \langle \mathcal{O}_{\Phi} \rangle = \lim_{r \to 0} \frac{1}{\sqrt{g^{(0)}}}\frac{\partial }{\partial \phi^{(0)}}S_{SUGRA}.
\end{equation}
Here $g$ is defined through $ds^2 = \frac{L^2}{r^2}\left(dr^2 + g_{mn} dx^m dx^n\right)$, and $g^{(0)}$ is the leading $\mathcal{O}(r^0)$ term in a series expansion of $g$. Similarly, $\phi^{(0)}$ is the leading $\mathcal{O}(r^{4-\Delta})$ term (with $\Delta$ the dimension of $O_{\Phi}$) in a series expansion of $\Phi$.

At this point the prescription \eqref{E:OPhiVEV} is not well defined since generically \eqref{E:OPhiVEV} will give divergent results. Simply ignoring these divergences may lead to wrong Ward identities. One prescription for systematically dealing with these singularities is holographic renormalization \cite{deHaro:2000xn,Bianchi:2001kw} (see \cite{Skenderis:2002wp} for a review). This prescription seems to work well even for non asymptotically AdS backgrounds \cite{Aharony:2005zr}. It comprises of adding a boundary action $S_{ct}$ to $S_{SUGRA}$. $S_{ct}$ may be composed only of the boundary values of the field $\Phi$, and is constructed so as to precisely cancel the $r \to 0$ divergences which appear in $\eqref{E:OPhiVEV}$.

In the current setup, we are interested in the expectation value of the energy momentum tensor \cite{Balasubramanian:1999re},
\begin{align}
    \langle T_{mn} \rangle &= \lim_{r \to 0}\frac{2}{\sqrt{g}}\frac{\partial}{\partial g^{mn}} \left(S_{SUGRA}+S_{ct}\right)\\
        &= T^{BY}_{mn} + X_{mn}
\end{align}
where $T^{BY}_{mn}$ is the Brown York stress tensor, which is what we would obtain by varying the on-shell Gibbons Hawking term in the action
and simply discarding the singularities.
More precisely, we are interested in the contribution of the trailing string, or moving quark, to the energy momentum tensor to linear order in $\kappa_5^2$, i.e.
\begin{equation}
\label{E:HRofTijlambda}
    \langle \delta T_{mn} \rangle = \delta T^{BY}_{mn} + \delta X_{mn}.
\end{equation}
The purpose of this section is to show that $\delta X_{mn}$ is diagonal and to evaluate the off diagonal parts of $\langle \delta T_{mn} \rangle$ in terms of the near boundary value of the metric fluctuations.

Expanding the various scalar fields in a power series in $r$, we find
\begin{equation}
\label{E:expansionF}
	\phi^I(r) = r^{4-\Delta_I}(\phi^{I\,(0)}+r^2 \phi^{I\,(2)} + \ldots) + r^{\Delta_I}(\bar{\phi}^{I\,(0)} + \ldots)
\end{equation}
where $\Delta_I$ is the dimension of the operator dual to $\phi^I$. If $\Delta_I$ is an integer then additional logarithmic terms will appear in the series expansion for $\phi^I$. Fields dual to irrelevant operators have $\Delta_I > 4$. In that case the power series expansion in $r$ is divergent at the boundary and the sources $\phi^{I\,(0)}$ should be taken to be infinitesimal, so that they vanish on-shell \cite{Witten:1998qj,deHaro:2000xn}.
For clarity, we shall omit the index $I$, and reinsert it only when required.
The coefficients $\phi^{(n)}$ in \eqref{E:expansionF} can be related to $\phi^{(0)}$ by solving the equation of motion for $\phi$ order by order in $r$. The coefficients $\bar{\phi}^{(n)}$ are related to $\bar{\phi}^{(0)}$ in a similar manner. $\bar{\phi}^{(0)}$ and $\phi^{(0)}$ are determined through the boundary conditions. For the case at hand the response of the field to the string is such that the leading near boundary contributions vanish. Using the same notation as in \eqref{LinearizedFluctuations}
\begin{equation}
\label{E:LF2}
    \phi \to \phi + \delta\phi
\end{equation}
we have $\delta \phi^{(0)} = 0$. Also, in our construction the background value of the fields $\phi$ depend only on the radial coordinate, meaning that in the notation of \eqref{E:LF2} $\phi^{(0)}$ is a constant.
A similar analysis follows for a near boundary expansion of the metric
\begin{equation}
\label{E:expansionG}
	g_{mn} = g_{mn}^{(0)}+g_{mn}^{(2)} r^2 + \bar{g}_{mn}^{(0)}r^4 + \ldots.
\end{equation}
Consider now the boundary counterterm action $S_{ct}$. As discussed earlier it must be composed of the near boundary metric $\gamma_{mn}(\epsilon) = \frac{L^2}{\epsilon^2}g_{mn}(\epsilon)$ and fields $\phi(\epsilon)$, where $\epsilon$ is the radial coordinate which will be taken to zero at the end of the calculation. Writing the metric dependence explicitly, we find
\begin{equation}
\label{E:Sctexpansion}
	S_{ct} = \frac{1}{2\kappa_5^2}\int \sqrt{\gamma}L_{ct}d^4x = \frac{1}{2\kappa_5^2} \int\sqrt{\gamma}\left(L_{\Lambda} + R L_{R} + \gamma^{ij}L_{ij}+\mathcal{O}(\epsilon^4)\right)d^4x,
\end{equation}
where the integrals are over a surface which is a distance $r=\epsilon$ from the asymptotically AdS boundary.
In \eqref{E:Sctexpansion} we have used
\begin{align}
\label{E:gammaexpansion1}
	\gamma_{mn}&=\epsilon^{-2}&\sqrt{\gamma} &\sim \epsilon^{-4}&\gamma^{mn}&\sim \epsilon^2&R^{m}{}_{nkl}&\sim\epsilon^0\\
	R_{mn}&\sim\epsilon^0&R&\sim\epsilon^2&\square&\sim\epsilon^2
\label{E:gammaexpansion2}
\end{align}
and we've also rewritten any $\square \phi^I L_I$ expressions as $\gamma^{mn}\partial_n \phi^I \partial_m L_I$.

We wish to evaluate the finite terms in
\begin{align}
	\frac{2}{\sqrt{g^{(0)}}}\frac{\partial }{\partial g^{(0)\,mn}}\left(S_{ct}+S_b\right)
	&=
	\lim_{\epsilon \to 0} \frac{L^2}{\kappa_5^2\epsilon^2}\frac{1}{\sqrt{\gamma}}\frac{\partial }{\partial \gamma^{mn}}\left(S_{ct}+S_b\right)\\
	&= \lim_{\epsilon \to 0} \frac{L^2}{\kappa_5^2 \epsilon^2}\left(
		-\gamma_{mn}\left(L_{ct}+L_{b}\right)
		+2\frac{\partial L_{ct}}{\partial \gamma^{mn}}
	\right).
\label{E:dSctdG0}
\end{align}
where $S_b = \int \sqrt{\gamma} L_b d^4 x$ is the on-shell boundary action which gets contributions from the various scalar fields.

We start from the last expression in the parenthesis in \eqref{E:dSctdG0}. Using \eqref{E:Sctexpansion}, the leading contributions to $\frac{\partial L_{ct}}{\partial \gamma^{mn}}$ which come from the trailing string will look like
\begin{equation}
\label{E:dLctdG0}
	\epsilon^{-2}\left(\delta (R_{mn} L_R) + \delta L_{mn} + \ldots \right).
\end{equation}
The second term in \eqref{E:dLctdG0} vanishes since it must involve two derivatives in a direction tangent to the radial AdS direction. One of these will always act on a background field $\phi^{(0)}$ which depends only on the radial direction, giving $\partial_i \phi^{(0)} = 0$.

Consider the first term in \eqref{E:dLctdG0}. $L_R$ and $\delta L_R$ are polynomials of the scalar fields with $L_R, \delta L_R \sim \epsilon^n$ with $n \geq 0$. (The inequality for $n$ follows by considering \eqref{E:expansionF}. If $n<0$ then $L_R, \delta L_R \sim \epsilon^n$ would imply a non vanishing source term for a non-normalizable operator.) So we can expand $L_R = \sum_{s=0}L_R^{(s)} \epsilon^s$ and $\delta L_R = \sum_{s=0}\delta L_R^{(s)} \epsilon^s$. Similarly $R_{ij} = 0$ for $i \neq j$ because the background metric $g$ is independent of the $x_i$ directions and $\delta R_{mn} = \sum_{s=4}R_{mn}^{(s)} \epsilon^s$  since the boundary conditions on the response of the metric to the string are that the metric vanishes close to the boundary.\footnote{In \cite{Friess:2006fk} it was shown that even when the boundary conditions imply that the metric fluctuations vanish at the AdS boundary, the stringy source induces an $\mathcal{O}(r^3)$ $k$ independent term in the series expansion for the metric fluctuations. This $\mathcal{O}(r^3)$ contribution is apparently related to the infinite mass of the quark and corresponds to a delta function at the location of the quark when going to real space. Regardless, it does not appear in the $1/k$ expansion that we are considering here.} This brings us to the conclusion that $\delta (R_{ij} L_R) \sim \epsilon^4$ so the first term in \eqref{E:dLctdG0} will not contribute to the stress energy tensor through \eqref{E:Sctexpansion} and \eqref{E:dSctdG0}.

Finally we consider contributions coming from $\gamma_{mn} \left( L_{b}+L_{ct}\right)$.
Expanding
\[
    \gamma_{mn} = \frac{L^2}{\epsilon^2}\left(g_{mn}^{(0)}+g_{mn}^{(2)}\epsilon^2+\bar{g}_{mn}^{(0)}\epsilon^4 + \ldots \right)
\]
and $L_{ct} + L_b = L_{\Lambda}^{(0)} +L_{b}^{(0)} + \mathcal{O}(\epsilon^2)$, we see that any finite non diagonal contributions of $\frac{1}{\epsilon^4}\gamma_{ij}\left(L_{ct}+L_b\right)$ to $\langle T_{mn} \rangle$ can only come from $\bar{g}_{mn}^{(4)} \left( L_{\Lambda}^{(0)} + L_{b}^{(0)}\right)$. But such expressions will also contribute to a divergent $\mathcal{O}(\epsilon^{-4})$ term in
$\sqrt{\gamma} \left(L_{\Lambda} + L_{b}\right)$ which will not be compensated by the Gibbons Hawking term since it involves the scalar fields. Therefore such a contribution will not exist.

We are left with evaluating the Brown York stress tensor.
In a general setting, this tensor is defined on a hypersurface $\Sigma$ whose outward pointing normal is $N_\mu$.  (Outward pointing means pointing toward the conformal boundary.)  The definition is
 \eqn{BYtensor}{
  T^{BY}_{mn} = K_{mn} - \left( K + {3 \over L} \right)
   g^\Sigma_{mn} \,,
 }
where $g^\Sigma_{mn}$ is the induced metric on $\Sigma$ and $K_{mn}$ is its extrinsic curvature. This expression comes from varying the Gibbons Hawking term in terms of the boundary metric $\gamma^{\Sigma}_{mn}$. The indices $m$ and $n$ run from $0$ to $3$ (along $\Sigma$) while $\mu$ runs over all five dimensions of the asymptotically AdS spacetime, whose radius of curvature near the boundary is $L$.  In general,
 \eqn{GeneralExtrinsic}{
  g^\Sigma_{mn} = g_{mn} - N_m N_n \qquad
  K_{mn} = -g_m{}^k \nabla_k N_n \,,
 }
but in axial gauge, when $\Sigma$ is a surface at fixed $r$, one has the simpler expressions
 \eqn{SpecificExtrinsic}{
  g^\Sigma_{mn} = g_{mn} \qquad
  K_{mn} = {1 \over 2 \sqrt{G_{rr}}} \partial_r g_{mn} \,.
 }
The metric of the boundary theory and the holographic stress tensor must be computed in a coordinated way, because both are affected by conformal transformations.  A consistent prescription is
 \eqn{HolographicMS}{
  g^{(0)}_{mn} = \lim_{r \to 0} {1 \over \alpha^2} g_{mn} \qquad
   \langle T_{mn} \rangle = {1 \over \kappa_5^2} \lim_{r \to 0}
    \alpha^2 T_{mn}^{BY} \,.
 }
For the ansatz \eno{BHansatz}, one finds $g^{(0)} = \eta = \diag\{-1,1,1,1\}$ and
 \eqn{HolographicStress}{
  \langle \delta T_{mn} \rangle = \lim_{r \to 0} {2\alpha^4 \over L}
    (H_{mn} - \eta^{lk} H_{lk} \eta_{mn}) \,,
 }
where we have used \eno{deltaGH}.

\end{appendix}

\clearpage
\bibliographystyle{ssg}
\bibliography{diffuse}
\end{document}